\documentclass[]{jpsj2}

\def\beq{\begin{equation}}
\def\eeq{\end{equation}}

\title{Dynamics of the Density Matrix in Contact with a Thermal Bath and the Quantum Master Equation}

\author{%
Takashi Mori$^{1}\thanks{ E-mail address: mori@spin.phys.s.u-tokyo.ac.jp }$
and 
Seiji Miyashita$^{1,2}\thanks{ E-mail address: miya@spin.phys.s.u-tokyo.ac.jp }$
}

\inst{%
$^1$ Department of Physics, The University of Tokyo
7-3-1 Hongo, Bunkyo-ku, Tokyo 113-0033, Japan \\
$^2$ CREST, JST, 4-1-8 Honcho Kawaguchi, Saitama 332-0012, Japan
}

\recdate{\today}

\abst{%
We study the structure of the time evolution of the density matrix 
in contact with a thermal bath in a standard projection operator scheme. 
In general, the equation of motions of the density matrix with dissipative 
effects tends to lead any initial state into a steady state.
The reduced density matrix of the system in the steady state is obtained
by tracing out the degree of freedom of the thermal bath 
from the density matrix for the equilibrium state of the total system. 
This reduced matrix is modified by the interaction, 
and is different from that of the equilibrium of the system alone.
In a commonly used equation of motion of the reduced density matrix,
we have three terms, i.e., a term of the quantum mechanical evolution 
of the system density matrix,
a term of the non-Markov evolution due to memory effects,
and a term depending on the initial total density matrix.
We make clear roles of the three terms by explicit calculations of the
contributions of the terms to the steady state density matrix.
By making use of the role of each term, 
the properties of the commonly used quantum master equation are examined. 
For example, if we do not include the imaginary part of the second term, 
the quantum master equation leads the reduced density matrix into that of
the equilibrium of the system without modification. 
On the other hand, if we include the imaginary part, 
the steady state density matrix of the system satisfies the
master equation up to the second order of the interaction. 
This property indicates that, in the representation that diagonalizes the system Hamiltonian,
the leading order (the second order) terms of 
the off-diagonal elements 
of the steady state solution of the master equation 
agree with those of the density matrix of the equilibrium state. 
}

\kword{%
quantum master equation, projection operator, coarse-grained dynamics,
relaxation phenomena}

\begin{document}
\maketitle

\section{Introduction} 

Recently, it has become important to understand explicit quantum dynamics in many fields.
For example, in the so-called single molecular magnets, the quantum dynamics of magnetization
has been extensively studied.\cite{SMM0,SMM1,SMM2,SMM3,SMM4,V15,MFE,SMM7,SMM8,SMM9}
In these systems, the energy levels are discrete, 
and the quantum mechanical dynamics of a wave function under a sweep of external field is
described by the combination of adiabatic and non-adiabatic transitions, 
where the Landau-Zener formula plays an important role. 
In addition to the pure quantum mechanical dynamic,
effects due to contact with a thermal bath also give important contributions. 
It has been pointed out that the magnetization shows a plateau 
in the magnetization process under a swept magnetic field.\cite{Fe2,V15,MFE}
To understand this phenomenon, a kind of rate equation for the populations of 
energy levels\cite{V15} and a quantum master equation\cite{MFE} 
have been studied.
It turned out that the appearance of the plateau is a generic phenomenon 
in systems in which the state changes almost adiabatically 
with a small inflow of heat from a thermal bath.
Quantum master equation approaches have also been used for
various properties of the magnetization processes of single molecular magnets.
\cite{SMM7,SMM8} Recently, the coherent driven Rabi oscillation has been observed
in V$_{15}$,\cite{SMM9} where decoherence processes play important roles.
The importance of quantum mechanical dynamics has also been pointed out 
in electron motion under a time-dependent field\cite{ELDY1}, and in 
various operations related to quantum computing and others.
\cite{QC1,QC2,QC3,QC4}
In this system, dissipation due to contact with the environments 
also play important roles.

To describe relaxation phenomena in quantum dynamics, 
we may derive the equation of motion of the density matrix of the system using
a method of the projection operator.~\cite{Kubo,Louisell,QME0,QME1,QME2,QME3,QME4,QME5,
QME6,QME7,QME81,QME82,QME9,QME10,Saeki,QME11,QME12,QME13,STM2000,STasaki,STasaki2}  
In general, we derive the equation from the Bloch equation of the total system.
The total system consists of 
the system (S), thermal bath (B), and interaction between them (I), 
the Hamiltonians of
which are given by ${\cal H}_{\rm S}$, ${\cal H}_{\rm I}$, and ${\cal H}_{\rm B}$, 
respectively:
\beq
{\cal H}_{\rm T}={\cal H}_{\rm S}+\lambda{\cal H}_{\rm I}+{\cal H}_{\rm B},
\eeq
where $\lambda$ is a constant that controls the strength of the interaction.
The dynamics of the density matrix of the total system $\rho_{\rm T}$ 
is given by the Bloch equation
\beq
{i\hbar}{\partial\rho_{\rm T}\over\partial t} 
= \left[ {\cal H}_{\rm T}, \rho_{\rm T} \right].
\eeq
Generally, we assume that 
the contact with the thermal bath causes dissipation effects, and that 
the system would relax to the thermal equilibrium state.\cite{STasaki,STasaki2}
To describe such processes, we adopt an equation for the time evolution of 
a reduced density matrix in which the
degrees of freedom of the thermal bath are traced out.
In general, the equation leads an initial state into a steady state. 
This steady state should be compatible with the equilibrium of the total system.
The reduced density matrix of the system in the steady state is obtained from that 
of the total system by decimating the degrees of freedom of the thermal bath, 
and it must be the steady solution of the time evolution
of the reduced density matrix. 
Because of the interaction between the system and the thermal bath,
the reduced density matrix of the steady state is modified 
from that of the equilibrium of the system itself.

The equation obtained by the projection method consists of three terms: 
a term of the quantum dynamics due to the
system Hamiltonian, 
a term of the non-Markov evolution due to memory effects,
and a term depending on the initial total density matrix.
We study the contributions of these three terms to the steady state explicitly.
We point out that the third term plays an intrinsic role in the steady state, 
and that artificial choices for the third term, e.g., a product form of the
system part and thermal bath part, would not make the steady state simpler.

Moreover, we discuss the properties of the quantum master equation in the
context of the equation of motion of the reduced density matrix.
A conventional quantum master equation leads the reduced density matrix into that of
the equilibrium of the system itself, while 
the time evolution of the reduced density matrix leads
the reduced density matrix into
a modified one owing to the interaction with the thermal bath.
The modified one agrees with the density matrix of the system interacting with the bath 
in the leading order of interaction. 
Namely, the second order terms of the off-diagonal elements of 
the solution in the representation that diagonalizes 
the system Hamiltonian correctly reproduces 
those of the equilibrium density matrix of the system interacting with the bath.
We study the roles of the real and imaginary terms of the second term of the 
equation of motion of the reduced density matrix, and clarified the
difference in the steady state.

\section{Formulation}

\subsection{Equation of the density matrix}
First, let us briefly review a standard derivation of 
an equation for the reduced density matrix of the system $\rho_{\rm S}$ by tracing out 
the degree of freedom of the bath from the
density matrix of the total system $\rho_{\rm T}$ :~\cite{Kubo}
\beq
\rho_{\rm S} \equiv {\rm Tr}_{\rm B}\rho_{\rm T},
\eeq
where ${\rm Tr}_{\rm B}$ denotes the trace operation for the degree of freedom of the 
thermal bath.
Here, we adopt a projection operator ${\cal P}$ to the Hilbert space of the system as
\beq
{\cal P}\rho_{\rm T} =\rho_{\rm B}{\rm Tr}_{\rm B}\rho_{\rm T},
\eeq 
and 
\beq
{\cal P}'=1-{\cal P},
\eeq
where $\rho_{\rm B}$ is the equilibrium density matrix of the thermal bath.
The equation of motion of the reduced density matrix ${\cal P}\rho_{\rm T}$ is given by
the following three parts:
\beq
{\partial{\cal P}\rho_{\rm T}\over\partial t}={\cal P}i{\cal L}{\cal P}\rho_{\rm T}
+{\cal P}i{\cal L}\int_{t_0}^{t}e^{(t-\tau){\cal P}'i{\cal L}} {\cal P}'i{\cal L}{\cal P}\rho_{\rm T}(\tau)d\tau
+{\cal P}i{\cal L}e^{(t-t_0){\cal P}'i{\cal L}} {\cal P}'\rho_{\rm T}(t_0),
\label{eq:LT}
\eeq
where 
\beq
i{\cal L}\rho_{\rm T}\equiv{1\over i\hbar}
\left[
{\cal H}_{\rm S}+\lambda{\cal H}_{\rm I}+{\cal H}_{\rm B}, \rho_{\rm T}
\right].
\eeq
This equation consists of three terms:
the first term denotes the quantum dynamics due to the 
system Hamiltonian, the second term represents
the non-Markov evolution due to memory effects, and 
the last term depends on the initial total density matrix $\rho_{\rm T}(t_0)$.
Here it should be noted that
this equation is part of the exact equations for the total density matrix 
together with the equation of ${\cal P}'\rho_{\rm T}$.
Thus, we cannot regard this relation as an equation of motion of $\rho_{\rm S}$, but
 a relation concerning $\rho_{\rm S}$ in terms of $\rho_{\rm T}(t_0)$.

Here, we study cases with a weak coupling, and consider the dynamics 
up to the second order of $\lambda$. Then eq. (\ref{eq:LT}) 
is expressed in the form (see Appendix A)
$$
{\partial\rho_{\rm S}\over\partial t}
={1\over{i\hbar}}\left[{\cal H}_{\rm S},\rho_{\rm S}\right]
+{\rm Tr}_{\rm B}i{\cal L}_{\rm I}\int_{t_0}^{t}e^{(t-\tau)i{\cal L}_0}i{\cal L}_{\rm I}
\rho_{\rm B}\rho_{\rm S}(\tau)d\tau
$$
\beq
+\lambda{\rm Tr_B}i{\cal L}_{\rm I}e^{(t-t_0)i{\cal L}_0}{\cal P}'\rho_{\rm T} (t_0) 
+ \lambda^2{\rm Tr_B}i{\cal L}_{\rm I}e^{(t-t_0)i{\cal L}_0}\int_0^1 (t-t_0)dx 
\: {\cal P}'e^{-x(t-t_0)i{\cal L}_0}i{\cal L}_{\rm I}{\cal P}'e^{x(t-t_0)i{\cal L}_0}{\cal P}'\rho_{\rm T}(t_0),
\label{eq:LT2}
\eeq
where
\beq
{\cal L}_0\rho_{\rm T} \equiv {1\over i\hbar}[{\cal H}_0,\rho_{\rm T} ] , \quad
{\rm and}\quad
{\cal L}_{\rm I}\rho_{\rm T} \equiv {1\over i\hbar}[{\cal H}_{\rm I},\rho_{\rm T} ] .
\eeq
Now, for the interaction, we adopt the form
\beq
{\cal H}_{\rm I}=XY,
\label{HIXY}
\eeq
where $X$ is an operator of the system and $Y$ is an operator of the thermal bath.
It is straightforward to extend eq. (\ref{HIXY}) into a more general form consisting
of the sum of interactions $(\sum_iX_iY_i)$. 

By the standard way, we derive an explicit form of the equation as
$${\partial\rho_{\rm S}\over\partial t}=
{1\over{i\hbar}}\left[{\cal H}_{\rm S},\rho_{\rm S}\right]$$
$$+\left({\lambda\over i\hbar}\right)^2\int_{0}^{t-t_0}
\left( 
 Xe^{-iu{\cal H}_{\rm S} /\hbar}X\rho_{\rm S}(t-u) e^{iu{\cal H}_{\rm S}/\hbar}\Psi(u)
-Xe^{-iu{\cal H}_{\rm S} /\hbar}\rho_{\rm S}(t-u) Xe^{iu{\cal H}_{\rm S}/\hbar}\Psi(-u) \right.$$
$$\left. 
-e^{-iu{\cal H}_{\rm S}/\hbar}X \rho_{\rm S}(t-u) e^{iu{\cal H}_{\rm S}/\hbar}X\Psi(u)
+e^{-iu{\cal H}_{\rm S}/\hbar}\rho_{\rm S}(t-u)X e^{iu{\cal H}_{\rm S} /\hbar}X\Psi(-u) \right)du
$$
\beq
+\lambda{\rm Tr_B}i{\cal L}_{\rm I}e^{(t-t_0)i{\cal L}_0}{\cal P}'\rho_{\rm T} (t_0) 
+ \lambda^2{\rm Tr_B}i{\cal L}_{\rm I}e^{(t-t_0)i{\cal L}_0}\int_0^1 (t-t_0)dx 
\: {\cal P}'e^{-x(t-t_0)i{\cal L}_0}i{\cal L}_{\rm I}{\cal P}'e^{x(t-t_0)i{\cal L}_0}{\cal P}'\rho_{\rm T}(t_0).
\label{QME2TC}
\eeq
Here,
\beq
X(t)=e^{it{\cal H}_{\rm S} /\hbar}Xe^{-it{\cal H}_{\rm S}/\hbar}.
\eeq
Because the second term on the right-hand side has the coefficient $\lambda^2$, 
we used the replacement\cite{QME81}
\beq
\rho_{\rm S}(t)=e^{-(t-\tau)i{\cal H}_{\rm S}/\hbar}\rho_{\rm S}(\tau)
 e^{(t-\tau)i{\cal H}_{\rm S}/\hbar}.
\eeq
Then, we have the following time-convolutionless form:
$${\partial\rho_{\rm S}\over\partial t}=L^{(2)}(\rho_{\rm S})\equiv {1\over{i\hbar}}\left[{\cal H}_{\rm S},\rho_{\rm S}\right]$$
$$+\left({\lambda\over i\hbar}\right)^2{\rm Tr}_{\rm B}\int_{0}^{t-t_0}
\left( XX(-u)\rho_{\rm S}(t)\Psi(u)-X\rho_{\rm S}(t)X(-u)\Psi(-u)\right)$$
$$\left. -X(-u)\rho_{\rm S}(t)X\Psi(u)+\rho_{\rm S}(t)X(-u)X\Psi(-u)\right)du
$$
\beq
+\lambda{\rm Tr_B}i{\cal L}_{\rm I}e^{(t-t_0)i{\cal L}_0}{\cal P}'\rho_{\rm T} (t_0) 
+ \lambda^2{\rm Tr_B}i{\cal L}_{\rm I}e^{(t-t_0)i{\cal L}_0}\int_0^1 (t-t_0)dx 
\: {\cal P}'e^{-x(t-t_0)i{\cal L}_0}i{\cal L}_{\rm I}{\cal P}'e^{x(t-t_0)i{\cal L}_0}{\cal P}'\rho_{\rm T}(t_0).
\label{QME2}
\eeq

$\Psi(t)$ is the autocorrelation function of $Y$ in the thermal bath expressed as
\beq
\Psi(t)={\rm Tr}_{\rm B}e^{it{\cal H}_{\rm B}/\hbar}Ye^{-it{\cal H}_{\rm B}/\hbar}Ye^{-\beta{\cal H}_{\rm B}}/Z_{\rm B}
=\langle Y(t)Y \rangle_{\rm B},
\eeq
where
\beq
Z_{\rm B}={\rm Tr}_{\rm B}e^{-\beta{\cal H}_{\rm B}},
\eeq
and $\langle \cdots \rangle_{\rm B}$ denotes the average in the equilibrium state of the thermal bath.

As natural properties of the thermal bath,
we assume the following properties of the correlation
\beq
\lim_{t\rightarrow\infty}\Psi(t)=0,
\label{P0}
\eeq
and the Kubo-Martin-Schwinger relation
\beq
\Psi (t) = \Psi (-t-i\hbar\beta ).
\label{KMS}
\eeq

Because the equilibrium state of the total system expressed as 
\beq
\rho_{\rm T}^{\rm eq}={e^{-\beta{\cal H}_{\rm T}}\over Z}, 
\quad Z={\rm Tr}e^{-\beta{\cal H}_{\rm T}}
\eeq
is the steady state of the time evolution
\beq
{\partial\rho_{\rm T}^{\rm eq}\over\partial t}={1\over i\hbar}\left[{\cal H}_{\rm T}, e^{-\beta{\cal H}_{\rm T}}/Z\right]=0,
\label{commute}
\eeq
the reduced density matrix  of the system
\beq
\rho_{\rm S}^{\rm eq}\equiv {\rm Tr}_{\rm B}\rho_{\rm T}^{\rm eq}.  
\eeq
must be the steady state of the time evolution (\ref{QME2}). 
In the following, we study the contributions to the time evolution of $\rho_{\rm S}^{\rm eq}$ 
from the three parts up to the second order of $\lambda$, 
and explicitly study how eq.(\ref{commute}) holds.

Here, let us briefly study the relation between the present way of discussion and the 
time-convolutionless (TCL) formalism. 
In order to study non-Markov noise, the 
TCL formalism has been introduced.\cite{QME7,QME81}
The time-convolution term in eq. (\ref{eq:LT2}) was transformed
into the TCL form
\beq
{\partial\over\partial t}\rho^{(2)}_{\rm S} (t)=i{\cal P}{\cal L}\rho^{(2)}_{\rm S} (t)
-\Psi(t)\rho^{(2)}_{\rm S}(t)
+i{\cal P}{\cal L}\theta(t)e^{i{\cal P}'{\cal L} t}{\cal P}' \rho_{\rm T}(0),
\eeq
where
\beq
\Psi(t)=i\left\langle {\cal L}_I {{\cal P}' \{S(t)R(-t)-1\}\over 1+{\cal P}' \{S(t)R(-t)-1\}} \right\rangle_{\rm B}
\eeq
with
\beq
R(t)=\exp_{\leftarrow}\left[i\int_0^td\tau e^{-i{\cal L}_0\tau}{\cal L}_I e^{i{\cal L}_0\tau}\right]
\eeq
and 
\beq
S(t)=\exp_{\rightarrow}\left[i\int_0^td\tau {\cal P}' e^{i{\cal L}_0\tau}{\cal L}_I e^{-i{\cal L}_0\tau}{\cal P}'\right].
\eeq
Here, $\exp_{\leftarrow}$ and $\exp_{\rightarrow}$ denote the time-ordered exponential functions.
Up to the second order of $\lambda$, this formula gives eq. (\ref{QME2}), which
has the TCL form. Thus, the argument in the present paper is for the TCL form. 
If we study the time-convolution form eq. (\ref{QME2TC}), we have a difference on the order of $\lambda^4$.
Up to the order of $\lambda^2$, they have the same form. Therefore, the present argument up to 
the order of $\lambda^2$ is good for both cases. However, as we will discuss in the next section, 
the solution of the quantum master equation has ambiguity on the order of $\lambda^2$.
Thus, the choice of higher-order terms gives different solutions. This point will be studied 
in the next section.

\subsection{Reduced density matrix for the equilibrium state}

First, we obtain the reduced density matrix of the system for the equilibrium state
up to the second order of $\lambda$.
We expand $\rho_{\rm T}^{\rm eq}$ as
$$
\rho_{\rm T}^{\rm eq}= {e^{-\beta({\cal H}_{\rm S}+{\cal H}_{\rm B})}\over Z'}\left(
1-\beta\int_0^1 dx 
e^{x\beta({\cal H}_{\rm S}+{\cal H}_{\rm B})} {\cal H}_{\rm I}e^{-x\beta({\cal H}_{\rm S}+{\cal H}_{\rm B})} \right.
$$
\beq
\left. 
+{\beta^2}\int_0^1 dx \int_0^x dy
e^{x\beta({\cal H}_{\rm S}+{\cal H}_{\rm B})} {\cal H}_{\rm I}e^{-x\beta({\cal H}_{\rm S}+{\cal H}_{\rm B})} 
e^{y\beta({\cal H}_{\rm S}+{\cal H}_{\rm B})} {\cal H}_{\rm I}e^{-y\beta({\cal H}_{\rm S}+{\cal H}_{\rm B})} 
+O(\lambda^3)\right),
\eeq
where $Z'=Z_{S}Z_{\rm B}-\lambda a+\lambda^2 b$ with 
$Z_{\rm S}={\rm Tr}_{\rm S}e^{-\beta {\cal H}_{\rm S}}$, 
$Z_{\rm B}={\rm Tr}_{\rm B}e^{-\beta {\cal H}_{\rm B}}$, 
\beq
a={\rm Tr}\beta\int_0^1 dx 
e^{x\beta({\cal H}_{\rm S}+{\cal H}_{\rm B})} {\cal H}_{\rm I}e^{-x\beta({\cal H}_{\rm S}+{\cal H}_{\rm B})}, 
\eeq
and 
\beq
b={\rm Tr} {\beta^2}\int_0^1 dx\int_0^x dy
e^{x\beta({\cal H}_{\rm S}+{\cal H}_{\rm B})} {\cal H}_{\rm I}e^{-x\beta({\cal H}_{\rm S}+{\cal H}_{\rm B})} 
e^{y\beta({\cal H}_{\rm S}+{\cal H}_{\rm B})} {\cal H}_{\rm I}e^{-y\beta({\cal H}_{\rm S}+{\cal H}_{\rm B})} ,
\eeq
where ${\rm Tr}_{\rm S} $ is the trace operation of the degree of freedom of the system.
We may include the average of the interaction ${\rm Tr}{\cal H}_{\rm I}e^{-\beta{\cal H}_{\rm B}}/Z_{\rm B}$ into
the system Hamiltonian, i.e., 
${\cal H}_{\rm S} \rightarrow 
{\cal H}_{\rm S}+{\rm Tr}_{\rm B}{\cal H}_{\rm I}e^{-\beta{\cal H}_{\rm B}}/Z_{\rm B}$, and then 
we can generally assume that  
\beq
{\rm Tr}{\cal H}_{\rm I}e^{-\beta{\cal H}_{\rm B}}=0.
\label{eq:av_H_I}
\eeq 
In this case,
\beq
a=0,
\eeq
and thus, up to the second order of $\lambda$, the $\rho^{\rm eq}_{\rm T}$ is
expressed in the following form: 
$$\rho_{\rm T}^{\rm eq} = {e^{-\beta({\cal H}_{\rm S}+{\cal H}_{\rm B})}\over Z_{\rm S}Z_{\rm B}}\left(
1-\lambda^2b-
\beta\lambda\int_0^1 dx 
e^{x\beta({\cal H}_{\rm S}+{\cal H}_{\rm B})} {\cal H}_{\rm I}e^{-x\beta({\cal H}_{\rm S}+{\cal H}_{\rm B})} \right.
$$
\beq
\left. 
+{\beta^2}\lambda^2\int_0^1 dx\int_0^x dy
e^{x\beta({\cal H}_{\rm S}+{\cal H}_{\rm B})} {\cal H}_{\rm I}e^{-x\beta({\cal H}_{\rm S}+{\cal H}_{\rm B})} 
e^{y\beta({\cal H}_{\rm S}+{\cal H}_{\rm B})} {\cal H}_{\rm I}e^{-y\beta({\cal H}_{\rm S}+{\cal H}_{\rm B})} 
\right) +O(\lambda^3)).
\label{rho2order}
\eeq

\subsection{Steady state}

Now, we substitute eq. (\ref{rho2order}) into eq. (\ref{QME2}).
The terms on the order of $O(\lambda^2)$ from the first, second, and third terms on the right-hand side 
of eq. (\ref{QME2}) are given by
\beq
\Delta_1\equiv {\beta\lambda^2\over i\hbar}
{e^{-\beta{\cal H}_{\rm S}}\over Z_{\rm S}}\int_0^1 dx \int_0^x dy
\left[{\cal H}_{\rm S},
X(-i\hbar\beta x)X(-i\hbar\beta y) \right]\Psi\left(i\hbar\beta(x-y)\right),
\label{term1}
\eeq
$$\Delta_2\equiv
\left({\lambda\over i\hbar}\right)^2{1\over Z_{\rm S}}\int_{0}^{t-t_0}
\left( XX(-u) e^{-\beta{\cal H}_{\rm S}} \Psi(u)-X e^{-\beta{\cal H}_{\rm S}} X(-u)\Psi(-u)\right.
$$
\beq
\left. -X(-u) e^{-\beta{\cal H}_{\rm S}} X\Psi(u)+ e^{-\beta{\cal H}_{\rm S}} X(-u)X\Psi(-u)\right)du
\label{term2}
\eeq
and
\beq \Delta_3\equiv
{-\beta\lambda^2\over i\hbar Z_{\rm S}}\int_0^1 dx\left[X, e^{-\beta {\cal H}_{\rm S}}
X(-t+t_0-i\hbar\beta x)\right]\Psi(-t+t_0-i\hbar\beta x),
\label{term3}
\eeq
respectively (see Appendix A).

As we have proven in Appendix B, we find
\beq
\Delta_1+\Delta_2+\Delta_3=0.
\label{balance}
\eeq
Namely, up to the second order of $\lambda$, the equilibrium 
density matrix of the system $\rho_{\rm T}^{\rm eq(2)}$ satisfies
\beq
{\cal L}\rho_{\rm T}^{\rm eq(2)}=0.
\label{steady2}
\eeq
Thus, we explicitly confirmed that the steady state of the equation of motion, eq. (\ref{QME2}), is given by 
the equilibrium state of the system under the influence of the interaction with the thermal bath, 
and that it differs from the equilibrium state of the system alone:
\beq
\rho_{\rm S}^{\rm eq0}\equiv e^{-\beta{\cal H}_{\rm S}}/Z_{\rm S}. 
\label{sum0}
\eeq

It should be noted that 
the third term gives a non zero contribution $\Delta_3$ to maintain the modified steady state, eq. (\ref{steady2}). 
Often, the initial density matrix is given by a product of
the form $\rho_{\rm T}(0)=\rho_{\rm S}(0)\rho_{\rm B}$, where the contribution of the third term is zero.
In such cases, eq. (\ref{balance}) is not satisfied. 
This means that the steady state of the total system is not given by a product state. 
Whenever we obtain a reduced density matrix for the steady state of the system,
information on the total density matrix is lost, and then eq. (\ref{QME2}) holds no more.
This apparent discrepancy is attributed to the fact that eq. (\ref{QME2}) does not
have a closed form for $\rho_{\rm S}$.
As far as we are concerned with the reduced density matrix, we may say that
the density matrix relaxes to $\rho_{\rm S}^{\rm eq}$, but this does not mean 
the total density matrix relaxes to $\rho_{\rm T}=\rho_{\rm S}^{\rm eq}\rho_{\rm B}$.

We assumed that the autocorrelation function in the bath decays to zero
in a long time limit, i.e., eq. (\ref{P0}). Thus, in a long time compared to the 
relaxation time of the autocorrelation function, we may ignore the third term.
Then, eq. (\ref{QME2}) apparently has a closed form, and we 
may regard it as an equation of motion of the reduced density matrix.
Usually, we derive a quantum master equation in this limit.
However, we must be careful with the condition to ignore the third term.
For example, the master equation composed only of the first and second terms cannot
be applied to the time evolution within the relaxation time of the thermal bath.

\section{Master Equation for the Equilibrium}

Now, we consider relation between conventional quantum master equations and 
eq. (\ref{QME2}).
When we study a relaxation process for the equilibrium of the system (S), 
we usually use a kind of quantum master equation, which is a closed equation of
motion of the density matrix of the system.
In eq. (\ref{QME2}), the third term disappears in the limit 
$t_0\rightarrow -\infty$ because of eq. (\ref{P0}). 
We can regard the remaining equation as an equation of motion of $\rho_{\rm S}$.
In this way, we can derive a quantum master equation. 
For example, we used the following compact form of the
master equation to study several time dependences of the quantum state.\cite{STM2000} 
\beq
{\partial\rho_{\rm S}\over\partial t}=L^{(2)}_0(\rho_{\rm S})
\equiv {1\over{i\hbar}}\left[{\cal H}_{\rm S},\rho_{\rm S}\right]
-\left({\lambda\over \hbar}\right)^2
\left(XR\rho_{\rm S}(t)-R\rho_{\rm S}(t)X
-X\rho_{\rm S}(t)R^{\dagger}+\rho_{\rm S}(t)R^{\dagger}X\right).
\label{eq:QME0}
\eeq
Here, the matrix $R$ is defined by the matrix elements
given by
\beq
R_{lm}={1\over\hbar}X_{lm}\Psi\left(E_l-E_m\over\hbar\right)
={1\over\hbar}X_{lm}{I\left(E_l-E_m\over\hbar\right)- I\left(E_m-E_l\over\hbar\right)
\over e^{\beta(E_l-E_m)}-1},
\label{Rlm}
\eeq  
where we use the basis that diagonalizes the system Hamiltonian ${\cal H}_{\rm S}$
\beq
{\cal H}_{\rm S}|m\rangle=E_m|m\rangle.
\eeq
$I(\omega)$ is given by the form
\beq
I(\omega)=\theta(\omega)I_0(\omega)
\eeq
with a function $ I_0(\omega)$. 
For the arbitrary autocorrelation function $\Psi(t)$, which satisfies the
Kubo-Martin-Schwinger relation (\ref{KMS}), we can assign eq. (\ref{Rlm}).

It is easy to verify that the equilibrium of the system gives the steady state
of this equation:
\beq
L^{(2)}_0(\rho_{\rm S}^{\rm eq0})=0.
\label{steady20}
\eeq
In the steady state of the above master equation, 
no effects of the interaction with the thermal bath do appear, i.e., eq. (\ref{steady20}),
although they exist in eq. (\ref{steady2}).
This property (\ref{steady20}) originates from the fact that
we neglected the contribution of the principal value integral 
coming from the relation
\beq
\int_0^\infty e^{i\nu t}dt=\pi\delta(\nu)+{\rm P}{i\over \nu}.
\label{eq:pri}
\eeq    
in the evaluation of the second term of eq. (\ref{QME2}) when we derived eq. (\ref{eq:QME0}).
\cite{STM2000}

Now let us study the contribution of the principal value integral.
In the previous section, we have shown that $\rho_{\rm T}^{\rm eq(2)}$, which includes 
the effects of the interaction with the thermal bath, 
is the steady state of the time evolution eq. (\ref{QME2}). 
For it, the three contributions canceled out $ \Delta_1+\Delta_2+\Delta_3=0 $ and
eq. (\ref{steady2}) held.
Taking into account the property (\ref{P0}), 
the third term disappears $(\Delta_3=0)$ in the limit $t_0\rightarrow -\infty$. 
Thus, in this limit, we study a quantum master equation given by
$${\partial\rho_{\rm S}\over\partial t}=L^{(2)}_{\rm RG}(\rho_{\rm S})
\equiv {1\over{i\hbar}}\left[{\cal H}_{\rm S},\rho_{\rm S}\right]$$
$$+\left({\lambda\over i\hbar}\right)^2\int_{0}^{\infty}
\left( XX(-u)\rho_{\rm S}(t)\Psi(u)-X\rho_{\rm S}(t)X(-u)\Psi(-u)\right)$$
\beq
\left. -X(-u)\rho_{\rm S}(t)X\Psi(u)+\rho_{\rm S}(t)X(-u)X\Psi(-u)\right)du.
\label{QME2RG}
\eeq
Here, we have 
\beq
\Delta_1+\Delta_2=0.
\label{D12}
\eeq
The contribution of the second term $\Delta_2$ consists of two parts:
one from the term $\pi\delta(\nu)$, which we call ``the real part" ($\Delta_{\rm 2R}$), 
and one from the principal value integral, 
which we call ``the imaginary part" ($\Delta_{\rm 2I}$):
\beq
\Delta_2=\Delta_{\rm 2R}+\Delta_{\rm 2I}.
\eeq
In Appendix C, we show that
the contribution of the principal value integral $\Delta_{\rm 2I}$ is the same as $-\Delta_1$:
\beq
\Delta_1=-\Delta_{\rm 2I}
\eeq
up to the order $\lambda^2$. 
Because the equilibrium density matrix of the system itself commutes with $\cal{H}_{\rm S}$,
the extra dynamics of the first term of the time evolution equation (\ref{QME2}),
i.e., $\Delta_1$, is due to the interaction with the thermal bath.
This motion is canceled by $\Delta_{\rm 2I}$.

Thus, we conclude that 
the imaginary part 
gives the renormalization of Hamiltonian of the system.
This effect has been pointed out as the Lamb shift in the driven damped oscillator 
system.\cite{Louisell}
We can take into account the renormalization of the system Hamiltonian 
by adopting the contribution of the principal value integral.

Now, we consider types of master equations.
We often need a quantum master equation that causes the relaxation 
to the equilibrium of the system. In such cases, we can have a desirable equation 
by discarding the principal value part. On the other hand, if we need to discuss
the effect of the contact with the bath,
we have to choose the master equation including the imaginary part. 
We have to choose the types according to the purpose of the study. 

It should be noted that 
even if the equation of motion is correct up to the order of $\lambda^2$,
the diagonal part of the solution has ambiguity on the order of $\lambda^2$.
Let $\rho_{\rm st}$ be the solution of the master equation (\ref{QME2RG})
\beq
L^{(2)}_{\rm RG}(\rho_{\rm st})=0.
\eeq
If we add the arbitrary traceless diagonal operator $W$ to $\rho_{\rm st}$
in the form 
\beq
\rho'=\rho_{\rm st}+\lambda^2 W.\,\quad {\rm Tr}W=0,
\eeq
eq. (\ref{QME2RG}) still holds up to the order of $\lambda^2$, because
for any traceless diagonal operator $W$ 
\beq
\Delta_1={1\over i\hbar}\left[{\cal H}_{\rm S},W\right]=0,
\eeq
and the contribution of $\lambda^2 W$ to $\Delta_2$ is on the order of $\lambda^4$.
Therefore, although $\rho_{\rm S}^{\rm eq(2)}$ satisfies eq. (\ref{QME2RG}) 
up to the order of $\lambda^2$, it does not necessarily agree with $\rho_{\rm st}$
on the order of $\lambda^2$. 
The diagonal elements of $\rho_{\rm st}$ and $\rho_{\rm S}^{\rm eq(2)}$
coincide in the 0th order of $\lambda$, 
and their off-diagonal elements agree with each other 
on the order of $\lambda^2$, which is the leading order. 
The part of order of $\lambda^2$ of the diagonal elements is related to
the $O(\lambda^2)$ terms of off-diagonal elements. 
Therefore, the $O(\lambda^2)$ term of the steady solution of the equation depends on 
the $O(\lambda^4)$ term of the equation. The choice of TC (eq. (\ref{QME2TC})) or  
TCL (eq. (\ref{QME2})) gives different $O(\lambda^4)$ terms, and thus they give
different solutions on the order of $\lambda^2$.
We may compare $L^{(2)}_{\rm RG}(\rho_{\rm S}^{\rm eq(2)})$ with 
that obtained in the TC formalism (i.e., from eq. (\ref{QME2TC})). However,
the evaluation of the TC formalism requires integration and thus
it is difficult to estimate it explicitly.

Finally, let us consider the relationship between eqs. (\ref{QME2}) and (\ref{eq:QME0}).
Although the reduced density matrices such as $\rho_{\rm S}^{\rm eq0}$ and
$\rho_{\rm S}^{\rm eq(2)}$ are the solutions of the master equation,
any decoupled initial condition in the form $\rho_{\rm S}^{\rm eq0}\rho_{B}$ 
cannot satisfy eq. (\ref{QME2}) as we have discussed in the
previous section.
Thus, it is a difficult problem to relate the master equation (\ref{eq:QME0}) and
the original relation, eq. (\ref{QME2}). 
Here, we may say that in the master equation we adopt the same mechanism 
of relaxation as that in eq. (\ref{QME2}), but they are essentially different types of equations. 


\section{Summary and Discussion}

We have studied the dynamics of the coarse-grained density matrix
$\rho_{\rm S}$. 
The interaction between the system and the thermal bath
causes a modification from the equilibrium density matrix of the system itself.
We explicitly calculated the roles of the three terms of the equation of motion 
(\ref{QME2}) in the steady state. 
We pointed out that the third term expressing the contribution of
the initial condition plays an important role, and that we cannot ignore it in the study
of the stationary state.

It is often mentioned that the diagonal element of 
${\cal L}(\rho_{\rm S}^{\rm eq0})$ is zero up to the second order of $\lambda$. 
Although this is true, it does not help 
in the study of the steady (equilibrium) state of the time evolution.
As time goes, the time evolution given by eq. (\ref{QME2}) always brings 
the state to $\rho_{\rm S}^{\rm eq(2)}$, but not to $\rho_{\rm S}^{\rm eq0}$ 
($=e^{-\beta{\cal}_{\rm S}}/Z_{\rm S}$).
We also pointed out that eq. (\ref{QME2}), or more generally eq. (\ref{eq:LT}), 
cannot be regarded as an
equation of motion of $\rho_{\rm S}$, but it is part of the equations of motion for the 
total density matrix.

The relationship between eq. (\ref{QME2}) and a conventional master equation was also
discussed. Generally, a master equation is regarded as an equation obtained in the
limit of $t_0\rightarrow -\infty$ in eq. (\ref{QME2}). There,
the roles of the real part (delta function) and imaginary part 
(principal value integral) of eq. (\ref{eq:pri}) were clarified.
The former drives the density matrix to that of the equilibrium of the system itself, 
and the latter modifies the system Hamiltonian due to the contact with
the bath.
It should be noted that 
the latter effect is not due to the thermal average of the interaction part.
This average has been taken into account for the system Hamiltonian 
by eq. (\ref{eq:av_H_I}). 
Thus, the modification due to the principal value integral is a kind of renormalization 
due to a dynamical process such as the Lamb shift.

We also pointed out that
although $\rho_{\rm S}^{\rm eq(2)}$ satisfies eq. (\ref{QME2RG}) 
up to the order of $\lambda^2$, it does not necessarily agree with the solution of the
equation $\rho_{\rm st}$ on the order of $\lambda^2$.
What we can say is the following:
The diagonal elements of $\rho_{\rm st}$ and $\rho_{\rm S}^{\rm eq(2)}$
coincide on the 0th order of $\lambda$, and 
their off-diagonal elements agree with each other 
on the order of $\lambda^2$, which is the leading order. 

Although the facts that we have derived are rather trivial, we hope that 
the explicit classification
of the roles will clarify the structure of equations for the dissipative 
quantum dynamics.

\section*{Acknowledgments}

The authors would like to thank Professor Chikako Uchiyama, Dr. Mizuhiro Saeki, 
and also Dr. Keiji Saito for their valuable discussion.
This work was partially supported by a Grant-in-Aid for Scientific 
Research on Priority Areas ``Physics of new quantum phases in superclean materials"  
and also by the Next Generation Super Computer Project, 
Nanoscience Program of MEXT.

\appendix
\section{}

The term of order $O(\lambda ^2)$ from the first term on the right-hand side of eq. (\ref{QME2}) is
obtained by substituting the $O(\lambda ^2)$ term of eq. (\ref{rho2order}):
\beq
{1\over i\hbar}[ {\cal H}_{\rm S} , {\beta^2} \frac{e^{-\beta {\cal H}_{\rm S}}}{Z_{\rm S}} 
\frac{{\rm Tr_B}e^{-\beta {\cal H}_{\rm B}}}{Z_{\rm B}} \int_0^1 dx \int_0^x dy \:
e^{x\beta ({\cal H}_{\rm S}+{\cal H}_{\rm B})} {\cal H}_{\rm I}e^{-x\beta ({\cal H}_{\rm S}+{\cal H}_{\rm B})} 
e^{y\beta ({\cal H}_{\rm S}+{\cal H}_{\rm B})} {\cal H}_{\rm I}e^{-y\beta ({\cal H}_{\rm S}+{\cal H}_{\rm B})}] .
\eeq
If we substitute ${\cal H}_{\rm I}=XY$ (eq. (\ref{HIXY})), 
we directly obtain $\Delta _1$ in eq. (\ref{term1}).

Next, the term of order $O(\lambda ^2)$ from the second term of eq. (\ref{rho2order}), 
which gives 
$\Delta _2$ in eq. (\ref{term2}), is obtained by replacing $\rho _{\rm S}$ 
in the second term of eq. (\ref{QME2}) 
with $e^{-\beta {\cal H}_{\rm S}}/Z_{\rm S}$.

Finally, we consider the contribution of the third term.
The third term on the right-hand side of eq. (\ref{eq:LT}) is expressed by
\beq
\lambda {\rm Tr_B} i{\cal L}_{\rm I} e^{(t-t_0){\cal P}'i{\cal L}_0{\cal P}'+\lambda (t-t_0){\cal P}'i{\cal L}_{\rm I}{\cal P}'}{\cal P}'\rho_{\rm T}(t_0).
\label{eq:term3_A}
\eeq
Using the formula
\beq
e^{A+\lambda B} = e^A \left( 1+\lambda \int_0^1 dx \: e^{-Ax}Be^{Ax} \right) + O(\lambda ^2)
\eeq
for any set of operators A and B, 
the third term (eq. (\ref{eq:term3_A})) up to the second order of $\lambda$ 
is given by
\beq
\lambda {\rm Tr_B} i{\cal L}_{\rm I} e^{(t-t_0){\cal P}'i{\cal L}_0{\cal P}'} 
\left( 1 + \lambda \int_0^1 (t-t_0)dx \: e^{-x(t-t_0){\cal P}'i{\cal L}_0{\cal P}'}
{\cal P}'i{\cal L}_{\rm I}{\cal P}'e^{x(t-t_0){\cal P}'i{\cal L}_0{\cal P}'} \right) 
{\cal P}'\rho_{\rm T} (t_0) + O(\lambda ^3) .
\label{eq:term3_A2}
\eeq
Using the relations
\begin{equation}
\begin{array}{ll}
{\rm Tr_B}{\cal P} &= {\rm Tr_B} \\
{\cal P}'i{\cal L}_0{\cal P}' &= {\cal P}'i{\cal L}_0 \\
e^{(t-t_0){\cal P}'i{\cal L}_0} &= {\cal P}' e^{(t-t_0)i{\cal L}_0} \\
{\cal P}i{\cal L}_{\rm I}{\cal P}' &= {\cal P}i{\cal L}_{\rm I} \quad ({\rm from \: eq. (\ref{eq:av_H_I})}),
\end{array}
\end{equation}
eq. (\ref{eq:term3_A2}) becomes
\beq
\lambda {\rm Tr_B}i{\cal L}_{\rm I} e^{(t-t_0)i{\cal L}_0} \left( 1 + \lambda \int_0^1 (t-t_0)dx \: {\cal P}'e^{-x(t-t_0)i{\cal L}_0}{\cal P}'i{\cal L}_{\rm I}{\cal P}'e^{x(t-t_0)i{\cal L}_0} \right) {\cal P}'\rho_{\rm T} (t_0) + O(\lambda ^3) .
\label{eq:term3_A3}
\eeq
Because of the relation
\beq
{\cal P}'i{\cal L}_{\rm I}{\cal P}' = -{\cal P}i{\cal L}_{\rm I} + i{\cal L}_{\rm I}{\cal P}',
\eeq
for any $n$,
\beq
{\cal P}' {\cal L}_0^n{\cal P} = 0
\eeq
holds. Thus, eq. (\ref{eq:term3_A3}) becomes
\beq
\lambda{\rm Tr_B}i{\cal L}_{\rm I}e^{(t-t_0)i{\cal L}_0}{\cal P}'\rho_{\rm T} (t_0) 
+ \lambda^2{\rm Tr_B}i{\cal L}_{\rm I}e^{(t-t_0)i{\cal L}_0}\int_0^1 (t-t_0)dx 
\: {\cal P}'e^{-x(t-t_0)i{\cal L}_0}i{\cal L}_{\rm I}{\cal P}'e^{x(t-t_0)i{\cal L}_0}{\cal P}'\rho_{\rm T}(t_0) .
\label{eq:term3_A4}
\eeq
Now, we substitute $\rho_{\rm T}(t_0)$ of the form of eq. (\ref{rho2order}) into eq. (\ref{eq:term3_A4}).
In the second term, we take only the 0-th order of $\lambda$, and then 
the second term disappears because 
\beq
{\cal P}'e^{-\beta({\cal H}_{\rm S}+{\cal H}_{\rm B})}=0.
\eeq
In the first term, the term of the first order of $\lambda$ 
contributes to the $O(\lambda ^2)$ term as
\beq
-{\lambda ^2 \beta \over Z_{\rm S}Z_{\rm B}}{\rm Tr_B}i{\cal L}_{\rm I}
e^{(t-t_0)i{\cal L}_0}{\cal P}'e^{-\beta ({\cal H}_{\rm S}+{\cal H}_{\rm B})} \int_0^1 dx 
\: e^{x\beta ({\cal H}_{\rm S}+{\cal H}_{\rm B})}{\cal H}_{\rm I}e^{-x\beta ({\cal H}_{\rm S}+{\cal H}_{\rm B})} .
\label{eq:term3_A5}
\eeq
Since ${\cal P}e^{-\beta {\cal H}_{\rm B}}{\cal H}_{\rm I} = 0$, 
\beq
-{\lambda ^2 \beta \over Z_{\rm S}Z_{\rm B}}{\rm Tr_B}i{\cal L}_{\rm I}e^{(t-t_0)i{\cal L}_0}
e^{-\beta ({\cal H}_{\rm S}+{\cal H}_{\rm B})} \int_0^1 dx 
\: e^{x\beta ({\cal H}_{\rm S}+{\cal H}_{\rm B})}{\cal H}_{\rm I}e^{-x\beta ({\cal H}_{\rm S}+{\cal H}_{\rm B})} .
\label{eq:term3_A6}
\eeq
We substitute ${\cal H}_{\rm I} = XY$, and then eq. (\ref{eq:term3_A6}) becomes
\begin{eqnarray}
-{\lambda ^2 \beta \over i\hbar Z_{\rm S}Z_{\rm B}}{\rm Tr_B} 
[XY,e^{-\beta ({\cal H}_{\rm S}+{\cal H}_{\rm B})}\int_0^1 dx 
\: X(-t+t_0-i\hbar \beta x)Y(-t+t_0-i\hbar \beta x)] \nonumber \\
= -{\lambda ^2 \beta \over i\hbar}\int_0^1 dx \: [X , {e^{-\beta {\cal H}_{\rm S}} \over Z_{\rm S}}
X(-t+t_0-i\hbar \beta x)] {\rm Tr_B} \left( {e^{-\beta {\cal H}_{\rm B}} \over Z_{\rm B}}Y(-t+t_0-i\hbar \beta x)Y 
\right) .
\label{eq:term3_A7}
\end{eqnarray}
Noting that 
\beq
{\rm Tr}_{\rm B}{e^{-\beta {\cal H}_{\rm B}} \over Z_{\rm B}}Y(-t+t_0-i\hbar \beta x)Y 
= \Psi (-t+t_0 -i\hbar \beta x),
\eeq
eq. (\ref{eq:term3_A7}) corresponds to eq. eq. ((\ref{term3}).

\section{}

From now on, we derive
\beq
L^{(2)}(\rho _{\rm S}^{\rm eq(2)}) = \Delta _1 + \Delta _2 + \Delta _3 = 0 .
\label{eq:B0}
\eeq
In order to prove eq. (\ref{eq:B0}), first we prove
\beq
(\Delta _1 + \Delta _2 + \Delta _3)_{t=t_0} = 0 ,
\label{eq:B1}
\eeq
and
\beq
\frac{d}{dt} (\Delta _1 + \Delta _2 + \Delta _3) = 0 .
\label{eq:B2}
\eeq

At $t=t_0$,
\begin{eqnarray}
\Delta _2 &=& 0, \label{Delta2t0} \\
\Delta _3 &=& -\frac{\lambda ^2 \beta}{i\hbar Z_{\rm S}Z_{\rm B}}{\rm Tr_B} [{\cal H}_{\rm I}, 
e^{-\beta ({\cal H}_{\rm S}+{\cal H}_{\rm B})} \int_0^1 dx 
\: e^{x\beta ({\cal H}_{\rm S}+{\cal H}_{\rm B})}{\cal H}_{\rm I}e^{-x\beta ({\cal H}_{\rm S}+{\cal H}_{\rm B})}] .
\end{eqnarray}
Here, it should be noted that $\Delta _1$ is time-independent 
and $\Delta _3$ is equal to eq. (\ref{eq:term3_A6}).
Now we reform $\Delta _3$ using the relation
\beq
[ A , e^B \int_0^1 dx \: e^{-Bx}Ae^{Bx} ] + [ B , e^B \int_0^1 dx 
\int_0^x dy \: e^{-Bx}Ae^{Bx}e^{-By}Ae^{By} ] = 0,
\eeq
which is derived from the second order of $\lambda$ in the expansion of
\beq
[ \lambda A + B , e^{\lambda A + B} ] = 0,
\eeq
and we have
$$
\Delta _3(t=t_0) = \frac{\beta ^2\lambda ^2}{i\hbar Z_{\rm S}Z_{\rm B}} {\rm Tr_B} [ {\cal H}_{\rm S}, $$
$$ \quad e^{-\beta ({\cal H}_{\rm S}+{\cal H}_{\rm B})} \int_0^1 dx \int_0^x dy 
 \: e^{x\beta ({\cal H}_{\rm S}+{\cal H}_{\rm B})}{\cal H}_{\rm I}e^{-x\beta ({\cal H}_{\rm S}+{\cal H}_{\rm B})} 
e^{y\beta ({\cal H}_{\rm S}+{\cal H}_{\rm B})}{\cal H}_{\rm I}e^{-y\beta ({\cal H}_{\rm S}+{\cal H}_{\rm B})}] $$
\beq
= -\Delta _1 .
\eeq
Therefore, eq. (\ref{eq:B1}) is proved.

Next, with a straightforward calculation, we find that
\begin{eqnarray}
\frac{d}{dt}\Delta _3 &=& -\frac{\beta \lambda ^2}{i\hbar} \int_0^1 dx \: \frac{d}{dt} 
\left\{ [ X , \frac{e^{-\beta {\cal H}_{\rm S}}}{Z_{\rm S}}X(-t+t_0-i\hbar \beta x) ] \Psi (-t+t_0-i\hbar \beta x) \right\} \nonumber \\
&=& -\frac{\beta \lambda ^2}{i\hbar} \int_0^1 dx \: \frac{1}{i\hbar \beta}\frac{d}{dx} 
\left\{ [ X , \frac{e^{-\beta {\cal H}_{\rm S}}}{Z_{\rm S}}X(-t+t_0-i\hbar \beta x) ] \Psi (-t+t_0-i\hbar \beta x) \right\} \nonumber \\
&=& -\left( \frac{\lambda}{i\hbar} \right) ^2 \frac{1}{Z_{\rm S}} 
\left\{ [ X , X(-t+t_0)e^{-\beta {\cal H}_{\rm S}} ] \Psi (t-t_0) -[ X , e^{-\beta {\cal H}_{\rm S}}X(-t+t_0) ] \Psi (-t+t_0) \right\} \nonumber \\
&=& -\frac{d}{dt} \Delta _2 ,
\end{eqnarray}
where we used the KMS relation (\ref{KMS}).
Thus, we proved eq. (\ref{eq:B2}).

Until now, we assumed the form ${\cal H}_{\rm I} = XY$, 
but it is not necessary to restrict this form of ${\cal H}_{\rm I}$ to show eq. ({\ref{eq:B0}).

\section{}

In this appendix, we prove that the principal value integral gives the $\lambda^2$ term in the
steady state of the time evolution equation (\ref{QME2}) when $t_0=-\infty$, 
which is the case of the quantum master equation in the reference.\cite{STM2000}
We put 
\beq
{\hat \Psi}(\omega ) = \frac{1}{2\pi} \int_{-\infty}^{\infty} dt \: \Psi (t) e^{-i\omega t}
\label{eq:psi_ft}
\eeq
in the second term of eq. (\ref{QME2}) and use eq. (\ref{eq:pri}). 
The contribution of the principal value integral is
$$\frac{\lambda ^2}{i\hbar ^2} {\rm P} \int_{-\infty}^{\infty} d\omega 
\: \sum_{l,m} \left[ \frac{{\hat \Psi}(\omega ) }{\omega - \omega _{lm}}X_{kl}X_{lm}\rho _{mn}(t)
+\frac{{\hat \Psi}(\omega ) }{\omega - \omega _{nm}}X_{kl}\rho _{lm}(t)X_{mn} \right. $$
\beq
\left. -\frac{{\hat \Psi}(\omega ) }{\omega - \omega _{lm}}\rho _{km}(t)X_{ml}X_{ln}
-\frac{{\hat \Psi}(\omega ) }{\omega - \omega _{kl}}X_{kl}\rho _{lm}(t)X_{mn} \right] ,
\label{eq:C_pvi}
\eeq
where 
\beq
\omega _{lm} \equiv \frac{E_l-E_m}{\hbar} .
\eeq
If we substitute $\rho _{\rm S}(t) = e^{-\beta {\cal H}_{\rm S}}/Z_{\rm S}$, eq. (\ref{eq:C_pvi}) becomes
$$\frac{\lambda ^2}{i\hbar ^2 Z_{\rm S}} {\rm P} \int_{-\infty}^{\infty} d\omega \: \sum_l 
\left[ \frac{{\hat \Psi}(\omega ) }{\omega - \omega _{ln}}X_{kl}X_{ln}e^{-\beta E_n} 
+ \frac{{\hat \Psi}(\omega ) }{\omega - \omega _{nl}}X_{kl}X_{ln}e^{-\beta E_l} \right. $$
\beq
\left. -\frac{{\hat \Psi}(\omega ) }{\omega - \omega _{lk}}X_{kl}X_{ln}e^{-\beta E_k} 
-\frac{{\hat \Psi}(\omega ) }{\omega - \omega _{kl}}X_{kl}X_{ln}e^{-\beta E_l} \right] .
\label{eq:C_sim}
\eeq
From eq. (\ref{eq:psi_ft}), we obtain
\beq
{\rm P} \int_{-\infty}^{\infty} d\omega \: \frac{{\hat \Psi}(\omega ) }{\omega -\omega _{ln}} = \frac{1}{2\pi} 
\int_{-\infty}^{\infty} dt 
\: \Psi (t) {\rm P} \int_{-\infty}^{\infty} d\omega \: \frac{e^{-i\omega t}}{\omega -\omega _{ln}} .
\eeq
Since
\beq
{\rm P} \int_{-\infty}^{\infty} d\omega \: \frac{e^{-i\omega t}}{\omega} = \left\{
\begin{array}{l}
-i\pi \quad (t>0) \\
i\pi \quad (t<0)
\end{array}
\right. ,
\eeq
\begin{eqnarray}
{\rm P} \int_{-\infty}^{\infty} d\omega 
\: \frac{{\hat \Psi}(\omega ) }{\omega -\omega _{ln}} &=& \frac{1}{2\pi} 
\int_{-\infty}^0 dt \: \Psi (t)e^{-i\omega _{ln}t} \cdot i\pi
+\frac{1}{2\pi} \int_0^{\infty} dt \: \Psi (t)e^{-i\omega _{ln}t} \cdot (-i\pi ) \nonumber \\
&=& -\frac{i}{2} \int_0^{\infty} dt \: \left( \Psi (t)e^{-i\omega _{ln}t} 
-\Psi (t-i\hbar\beta )e^{i\omega _{ln}t} \right) .
\end{eqnarray}
Therefore, the first term of eq. (\ref{eq:C_sim}) is 
\beq
A_1=
-\frac{\lambda ^2}{2\hbar ^2 Z_{\rm S}} \int_0^{\infty}dt \sum_l 
(\Psi (t)e^{-i\omega _{ln}t} -\Psi (t-i\hbar\beta )e^{i\omega _{ln}t})X_{kl}X_{ln}e^{-\beta E_n} .
\label{eq:C_term1}
\eeq
If we assume eq. (\ref{P0}), we obtain
\beq
\int_0^{\infty} dt \: \Psi (t-i\hbar\beta )e^{i\omega _{ln}t} = \int_0^{\infty} dt \: \Psi (t)
e^{i\omega _{ln}t}e^{-\beta\hbar\omega _{ln}}
+i\hbar\beta \int_0^1 dx \: \Psi (-i\hbar\beta x)e^{\beta\hbar\omega _{ln}x}e^{-\beta\hbar\omega _{ln}}
\eeq
by changing the path of integral.
Therefore,
$$A_1 = -\frac{\lambda ^2}{2\hbar ^2 Z_{\rm S}} \int_0^\infty dt \sum_l \left\{ \Psi (t)e^{-i\omega _{ln}t}X_{kl}X_{ln}e^{-\beta E_n} -\Psi (t)e^{-i\omega _{nl}t}X_{kl}X_{ln}e^{-\beta E_l} \right\}$$
\beq
+\frac{i\beta\lambda ^2}{2\hbar Z_{\rm S}} \int_0^1 dx \sum_l \Psi (-i\hbar\beta x)e^{\beta (E_l-E_n)x}X_{kl}X_{ln}e^{-\beta E_l} .
\eeq
Similarly, the second term of eq. (\ref{eq:C_sim}) is expressed as
$$A_2=-\frac{\lambda ^2}{2\hbar ^2 Z_{\rm S}} \int_0^\infty dt \sum_l \left\{ \Psi (t)e^{-i\omega _{nl}t}X_{kl}X_{ln}e^{-\beta E_l} -\Psi (t)e^{-i\omega _{ln}t}X_{kl}X_{ln}e^{-\beta E_n} \right\}$$
\beq
+\frac{i\beta\lambda ^2}{2\hbar Z_{\rm S}} \int_0^1 dx \sum_l \Psi (-i\hbar\beta x)e^{\beta (E_n-E_k)x}X_{kl}X_{ln}e^{-\beta E_n} .
\eeq
Therefore, the contribution of the first and second terms of eq. (\ref{eq:C_sim}) is
\begin{eqnarray}
&A_1+A_2=\frac{i\beta\lambda ^2}{\hbar Z_{\rm S}} \int_0^1 dx \sum_l \Psi (-i\hbar\beta x)X_{kl}e^{-\beta E_l}e^{\beta E_lx}X_{ln}e^{-\beta E_nx} \nonumber \\
&= \frac{i\beta\lambda ^2}{\hbar Z_{\rm S}} \int_0^1 dx \left( Xe^{-\beta {\cal H}_{\rm S}}X(-i\hbar\beta x) \right) _{kn} \Psi (-i\hbar\beta x) .
\end{eqnarray}
Similarly, the contributions of the third and fourth terms of eq. (\ref{eq:C_sim}) are
\beq
A_3+A_4=-\frac{i\beta\lambda ^2}{\hbar Z_{\rm S}} \int_0^1 dx \left( e^{-\beta {\cal H}_{\rm S}}X(-i\hbar\beta x)X \right) _{kn} \Psi (-i\hbar\beta x) .
\eeq
After all, the contribution of the principal value integral is
\beq
A=A_1+A_2+A_3+A_4=
\frac{i\beta\lambda ^2}{\hbar Z_{\rm S}} \int_0^1 dx 
[ X , e^{-\beta {\cal H}_{\rm S}}X(-i\hbar\beta x) ] \Psi (-i\hbar\beta x) ,
\eeq
which is the same as $\Delta _3(t=t_0)$.

Because of eqs. (\ref{eq:B2}) and (\ref{Delta2t0}), 
we found that $\Delta _3(t=t_0) = -\Delta _1$.
Taking into account that $\Delta_1$ does not depend on $t_0$,
we conclude that the contribution of the principal value integral $A$ 
in the limit $t_0\rightarrow -\infty$ is the same as $-\Delta _1$.

\end{document}